\documentclass[twocolumn]{revtex4}
\usepackage{graphicx,amsmath}

\begin{document}

\title { Magnetic field asymmetry  of non-linear transport in Aharonov Bohm rings}
\author{L. Angers}
\affiliation{Laboratoire de Physique des Solides, Associ\'e au CNRS, B\^atiment 510, Universit\'e Paris-Sud, 91405, Orsay, France}
\author{E. Zakka-Bajjani}
\affiliation{Laboratoire de Physique des Solides, Associ\'e au CNRS, B\^atiment 510, Universit\'e Paris-Sud, 91405, Orsay, France}
\author{R. Deblock}
\affiliation{Laboratoire de Physique des Solides, Associ\'e au CNRS, B\^atiment 510, Universit\'e Paris-Sud, 91405, Orsay, France}
\author{S. Gu\'eron}
\affiliation{Laboratoire de Physique des Solides, Associ\'e au CNRS, B\^atiment 510, Universit\'e Paris-Sud, 91405, Orsay, France}
\author{A. Cavanna}
\affiliation{CNRS Laboratoire de Photonique et Nanostructures, CNRS,Route de Nozay, 91460 Marcoussis, France.}

\author{U. Gennser}
\affiliation{CNRS Laboratoire de Photonique et Nanostructures, CNRS,Route de Nozay, 91460 Marcoussis, France.}
\author{M. Polianski }\affiliation{D\'epartement de Physique Th\'eorique, Universit\'e de Gen\`eve, CH-1211 Gen\`eve 4, Switzerland}
\author{H. Bouchiat}
\affiliation{Laboratoire de Physique des Solides, Associ\'e au CNRS, B\^atiment 510, Universit\'e Paris-Sud, 91405, Orsay, France}

\begin{abstract}
Fundamental  Casimir-Onsager symmetry rules for linear response do not apply to non linear transport. This   motivates the investigation of  nonlinear dc  conductance of mesoscopic GaAs/GaAlAs rings in a 2 wire configuration. The second order  current response to a potential bias is of particular interest. It is  related to the sensitivity of conductance fluctuations to this bias and contains information on electron interactions not included in the linear response. In contrast with the linear response which is a symmetric function of magnetic field we find that this second order response exhibits a field dependence which contains an antisymmetric  part. We analyse the flux periodic and aperiodic components of this asymmetry and find that they only depend on the conductance of the rings which is varied by more than an order of magnitude. These  results are in good agreement with recent theoretical predictions relating this  asymmetric response  to the electron interactions.   
\end{abstract}

\maketitle

 Magneto-transport  properties of mesoscopic systems are sensitive to interferences between electronic waves  and exhibit  characteristic signatures of  phase coherence. \cite{intro} Among them are  universal conductance fluctuations (UCF) leading to reproducible sample specific magnetoresistance patterns,   which in a ring geometry are modulated by the flux periodic Aharonov-Bohm (AB) oscillations. Beside these effects on the linear conductance, it has been shown that  mesoscopic systems exhibit  rectifying properties related to the absence of spatial  inversion symmetry of the disorder  or confining potential. This gives rise  to a quadratic term in the $I,V$  relation: 
$I=G_1 V + G_2 V^2$. This non linearity  was predicted  theoretically \cite{altkm} and observed experimentally
 more than 15 years ago \cite{pionexp}. It was understood as a direct consequence of the sensitivity of conductance fluctuations to the Fermi energy  with a characteristic scale given by the Thouless energy, $E_c =  h/\tau_D$,  where $\tau_D= L^2/D$ is the diffusion time across the sample of size $L$. More recently it has been pointed out that non linear transport coefficients do not obey  Casimir Onsager symmetry  rules \cite{onsager} and may include, in a 2 wires measurement, a component antisymmetric  in magnetic field  with a linear  dependence at low field.  On the experimental side the existence of a term  linear both  in field and current  was found in  macroscopic helical structures  and attributed to magnetic self inductance effects \cite{rikenbis}. It was subsequently observed in carbon nanotubes  \cite{rikennano} and suggested to be related to  their helical structure. More generally, this field asymmetry     of non linear transport has been theoretically  shown to be related to electron-electron interactions both in   chaotic  \cite{sanchez} and diffusive \cite{spivak} systems. At the single impurity level, this effect  can be simply viewed as  the modification of the electron density $d n(\vec r)$ around a scatterer  in the presence of a current through the sample  \cite{landauer,zwerger}.  Due to Coulomb interactions this results,  in a modification of the potential  around the impurity by  a component linear in current.  In a phase coherent sample,  this bias induced change of disorder potential $d U_{dis}$ modifies the  conductance fluctuations giving rise to the nonlinear conductance $G_2$  defined as:  $ G(V)= G_1 +  G_2 V +..$.  Just like the chemical potential measured in a multiprobe transport measurement \cite{intro},   $d U_{dis}$  exhibits field dependent fluctuations which have both symmetric and antisymmetric parts, including a component linear in  B at low field. As a result $G_2$   has  a  symmetric component in $B$  and  an antisymmetric  one, respectively equal to  $G_2^{S,AS}= (G_2(B)\pm G_2(-B))/2$ which both vary on  the flux scale $\Phi_c$  characteristic of  conductance fluctuations.  The antisymmetric component only exists in the presence of electron-electron interactions. The typical amplitudes of these components  have been calculated in a 2D  system \cite{spivak, sanchez, polianski} and yield for weak interactions:
\begin{equation}
\delta G_2^S  \sim \delta G_1 (e/E_c), \,  \delta G_2^{AS} \sim  \gamma_{int}\frac{\delta G_1}{g}f(|\Phi/\Phi_c|)\delta G_2^S 
\label{astheo}
\end{equation}
where the function $f(x)$ is equal to  $x$  for $0<x\ll 1$ and to $1$ for $x\gg1$.  $g=<G_1> $   and $\delta G_1\simeq 1 $ are  the  average conductance and the typical amplitude 
 of $G_1$  fluctuations in units of $e^2/h$. In the weak interaction regime investigated in [8]
$\gamma_{int}=2\nu dU_{dis}(\vec r)/dn(\vec r)\ll 1$, where $\nu$ is the
density of states per unit surface. In the self-consistent treatment of Coulomb interaction
\cite {christen,polianski}, the interaction constant is determined by the ratio of typical
charging energy of the sample $\sim e^2/C$, and mean level spacing
$\Delta$ as  $\gamma_{int}=1/(1+C\Delta/2e^2)$ (the limit
$\gamma_{int}\ll 1$ of Eq. \ref{astheo} corresponds to $dU/dn=e^2/2C$)\cite{orderofmag}. The $1/g$ factor in Eq.\ref{astheo} indicates that  the field asymmetry should be easily detectable  in low conductance samples where $g$ does not exceed 10, but is not  observable in metallic mesoscopic systems   \cite{haussler}. Carbon nanotubes  \cite{cobden} are in principle good candidates, but  it is delicate to distinguish  the effects due to the tube helicity from mesoscopic ones.  Semiconducting samples are  well suited for such investigations since they combine rather  low conductance and large sensitivity to small fluxes \cite{lofgren, marcus, leturq}.  We have  measured the non linear conductance  of  2 terminal   GaAs/GaAlAs rings with only a few conducting channels. We find that $G_2$, like $G_1$, exhibits both AB oscillations and UCF conductance fluctuations.  We show evidence of the existence of  an asymmetry in magnetic field in $G_2$ from which   we deduce the amplitude of the  electron interactions.  
\begin{figure}
\begin{center} 
\includegraphics[width=7cm]{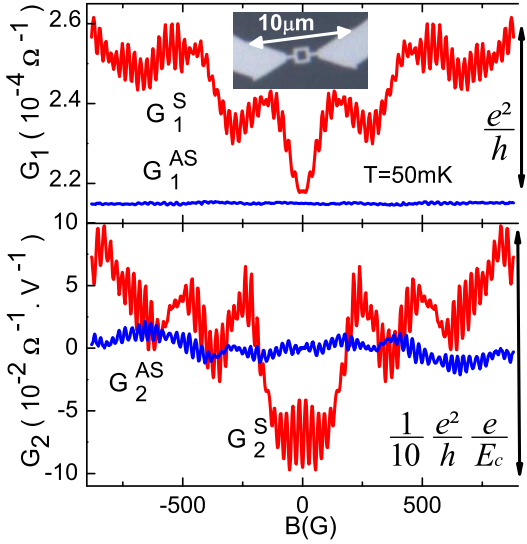}
\end{center}
\caption{Field dependence of $G_1 ^{S,AS}$  and  $G_2 ^{S,AS}$,($G_1 ^{AS}$  shifted).  These data were  obtained on  ring$_1$ in its initial state for a  current of 10nA. Inset: micrograph of  ring$_1$.}
\label{fig1} 
\end{figure}

The  square  rings   investigated in this experiment were obtained by shallow  etching through an aluminum mask, of a 2D electron gas (2DEG) of density $n_e=3.8 \times10^{15}\:$m$^{-2}$ at the interface of a GaAs/GaAlAs heterojunction  with  Si donors. Contrary to  previous experiments \cite{cobden, lofgren, marcus, leturq}, there is no electrostatic gate on the samples \cite{riken05}. Due to depletion effects the real width of the rings is smaller than the etched one and is determined  from  magneto-transport data.  We present data on 2 rings, of circumference $L=4.8\mu m$ and   respective width $W = 0.3$, $0.45\pm 0.05 \mu m$.  The elastic mean free path $l_e$ extracted  from  the  conductance at 4K varies  between 1 and $2\mu m$, which is less than the value of the initial 2DEG and comparable with  the  side of the square ring. In situ modifications of the samples were   obtained by  short illuminations with an electro-luminescent diode resulting in an  increase of  width and conductance of  the rings. It was  also possible to change the disorder configuration  by applying current pulses in the 10 to 50 $\mu A$ range which  decrease the average conductance of the samples.  We could therefore, with only 2 samples,  investigate a conductance  ranging  from $g=1$ to 20. The measurements were conducted  via  filtered   lines in a dilution refrigerator between 25mK and 1.2K.   The samples were biased with ac current of frequency $\omega/2\pi \simeq 30Hz$ in the  few nA range and  voltage was measured with a low noise amplifier followed  by lock-in amplifiers detecting the first and second harmonics  response $V_1 \cos \omega t  $ and $V_2 \cos 2\omega t$. The amplitude of the ac modulation was chosen   to maximize the second harmonics signal in the regime where it is still  quadratic with   the current modulation amplitude. In this regime the second order conductance $G_2 $ is simply related to  $V_2$  and $V_1$  by $G_2 = -2  V_2 I /V_1^3 $. As shown in  Fig.\ref{fig1} both  $G_1(B)$  and  $G_2(B)$ exhibit  $h/e$ periodic  AB oscillations modulated by  UCF fluctuations.

\begin{figure}
\begin{center} 
\includegraphics[width=8cm]{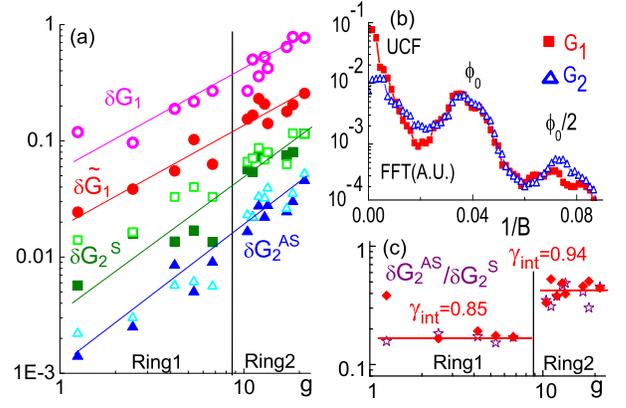}
\end{center}
\caption{(a)Amplitudes of the  UCF and AB components of $G_1$ in units of $e^2/h$, $ G_2^{S}$ and $ G_2^{AS}$ in $(\Omega V)^{-1} $ as a function of the  conductance of the rings. The open  and  closed symbols  correspond to  UCF and AB integrated peaks (straight lines are  guides to the eyes). (b)Fourier transforms of $G_2$ and $G_1$ renormalised so that the amplitude  of the AB peaks are identical. (c)Conductance dependence of the ratio $\delta G_2^{AS}/\delta G_2^{S}$ for UCF (open stars)  in comparison with Eq.\ref{compexptheo} (diamonds), where $\gamma_{int}=0.85$  for ring$_1$ and $0.94$ for ring$_2$. } 
\label{fig2} 
\end{figure}

More remarkable, whereas $G_1(B)$ is a symmetric function of magnetic field  as expected  in a 2 wires configuration, $G_2(B)$   exhibits a component antisymmetric in field $G_2^{AS}$. To compare the  field asymmetry of $G_2$ on the various samples (Fig.2), we extract the  amplitude of UCF and  AB oscillations from the Fourier transform  of $G_2^{S,AS}(B)$ and $G_1(B)$. The integrated  UCF and AB peaks are  noted $\delta G_2^{S,AS}$  and  $\delta\tilde  G_2^{S,AS}$  and similarly for $G_1$.   These averages  performed on a flux range much larger than $\Phi_c$ do  not depend on this  range.  We  find that all these quantities   are   aligned on logarithmic plots as a function of $g$. We  first note  the relatively large amplitudes of the AB oscillations  $\delta\tilde  G_2^{S,AS}$ which are of the order of the UCF components $\delta G_2^{S,AS}$  in contrast with the related quantities in  $G_1$. The decrease  of  $\delta G_1$ and $\delta\tilde G_1$ to values much below 1  at low  $g$  seems to be at odd with the universal character of conductance fluctuations  in $ G_1$ established  deep in the diffusive regime $g \gg 1$. However this universal regime is not expected  for $g\simeq 1$. The even larger  dependence measured in $\delta G_2^{S}(g)$ indicates, according  to  (\ref{astheo}), that $E_c = g\Delta$  decreases with $g$. This  finding can be attributed to the physics involved in the illumination process  which   increases the width of the rings   (resulting in a decrease of $\Delta$)  and also decreases the elastic scattering time. Finally  we find  that the ratio $\delta G_2^{AS}/\delta G_2^{S}$  depends only slightly on $g$ and is equal to $0.3\pm 0.2$(see  Fig.2c).  This result is  {\it a priori} in contradiction with Eq.\ref{astheo} from which a $1/g$ dependence of $\delta G_2^{AS}/\delta G_2^{S}$ is expected. However Eq.\ref{astheo} is only valid for $\gamma_{int} \ll 1$ in  partially closed  quantum dots, i.e. whose   classical resistance  is dominated by the contacts but still much lower than quantum resistance. It was recently found by one of us \cite{butpol2}  using random matrix theory  that  in these  systems    $\delta G_2^{S}$ also  strongly depends on  $\gamma_{int}$,  when it is not negligible compared to 1,  and in the limit $\Phi \gg \Phi_c$:
\begin{equation}
\delta G_2^{AS}/\delta G_2^{S}= 1 /\sqrt{1+2(g/\delta G_1)^2(1/\gamma_{int}-1)^2} 
\label{compexptheo}
\end{equation}
identical to (\ref{astheo}) in the limit of small $\gamma_{int}$.  It is interesting to note that Eq.\ref{compexptheo} predicts  $\delta G_2^{AS}/\delta G_2^{S}$  independent of $g$ in the limit  $\gamma_{int} =1$. It  describes remarkably well our experimental results, yielding $\gamma_{int}\simeq 0.94\pm 0.02$ for ring$_2$ and  $\gamma_{int}\simeq 0.85\pm 0.02$ for ring$_1$ . These values  are very close to the estimated values of $\gamma_{int}=1/(1+C\Delta/2e^2)\simeq 0.98 $  from the geometry of our samples. Our determination of the interaction constant from the dimension less quantity $\delta G_2^{AS}/\delta G_2^{S}$   is much more accurate than   the estimation done in reference \cite{marcus} from the analysis of the low field component of $G_2 ^{AS}$  only,  but consistent with it. Finally, our  results show that our rings are in a  highly interacting regime  and  behave as  partially closed quantum dots due to their relatively long mean free path.
\begin{figure}
\begin{center} 
\includegraphics[width=7cm]{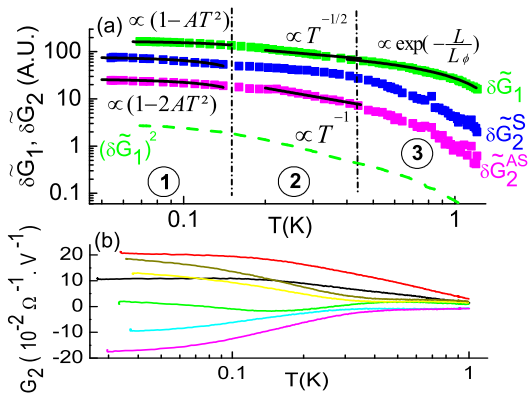}
\end{center}
\caption{(a)Temperature dependence of the  amplitude of the AB oscillations in $G_1$, $G_2^{S}$ and  $G_2^{AS}$ (ring$_2$). Continuous lines are the fits corresponding to the regions (1) $k_BT \ll E_c$  (2) $k_BT \gg E_c$ but $L \ll L_\phi$ (3) $L > L_\phi$.(b)Examples of temperature dependence of  $G_2$ in ring$_2$ for different values of magnetic field between -1000 and 1000 Gauss.}
\label{fig3}  
\end{figure}    
  
\begin{figure}
\begin{center} 
\includegraphics[width=7cm]{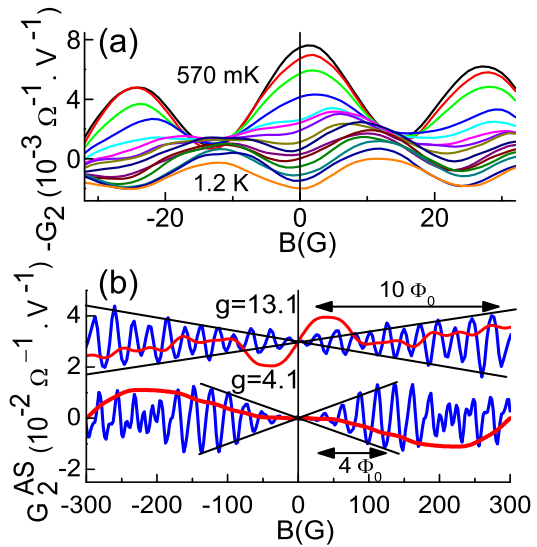}
\end{center}
\caption {(a)Low field dependence of $G_2$  on ring$_1$ for different temperatures.(b) Field dependence of $G_2^{AS}$  after low pass (light line) and high-pass (bold line)  filtering , on  ring$_1$ and   ring$_2$  respectively lower  and upper curves (shifted for clarity). Note the extinction of  AB oscillations in the vicinity of $B=0$.}
\label{fig4}
\end{figure}

   The temperature dependence of  $\delta\tilde G_1$ and  $\delta\tilde G_2^{S,AS}$ are shown  on  Fig.\ref{fig3}a. As  already observed \cite{washweb86} $\delta\tilde G_1$  is only weakly $T$ dependent below the Thouless energy like $1-A T^2 $  with $A =2(k_B/E_c)^2$ and decays at higher temperature  as  $T^{-1/2}$. Deviations   above $0.5K$ are consistent with  $\exp (-L/L_\phi(T))$ with a $T$ dependence of the  phase coherence length as $T^{-1/2}$ but this last fit  on a small range of temperature is not unique and  just indicative. $\delta\tilde G_2^{S,AS}$ have similar $T$ dependences, which are nearly identical   to $(\delta\tilde G_1(T))^2$ with a $T ^{-1}$ decay  in the limit $k_BT \gg E_c$ with $L\ll L_\phi$, in agreement with theoretical predictions \cite{spivak,polianski}. The temperature dependence of   $G_2$ at fixed  magnetic fields is depicted in  Fig.\ref{fig3}b. In the same way as observed  on $G_1$ \cite{spivak2},  $G_2(T)$ exhibits  a non monotonous  variation  with temperature on  the scale $E_c$ which randomly fluctuates  with magnetic field. In some cases we also  observed  (see  Fig.\ref{fig4}a) that the phase of the AB oscillations in $G_2$ depends on temperature.   Surprisingly  it remains pinned either to $0$ or $\pi$ at  zero field with  the appearance of a second harmonics contribution in the region of temperature where the sign change occurs. At larger field the phase  takes any value between $0$ and $\pi$. Note that this last effect is observed both in $G_1$ and $G_2$ as a result of phase modulation of the AB oscillations by the UCF, but these phase modulations are symmetric in field on $G_1$, and not on $G_2$.   In short, we find that AB oscillations  on $G_2$ are symmetric in zero field and the asymmetry only appears at higher field.  This  is also clearly seen in  the AB oscillations on $G_2^{AS}$. We observe in all samples that their amplitude vanish linearly at zero field (Fig.4b).

We now propose   a simple explanation of the extinction of AB oscillations at low field in $G_2 ^{AS}$ linked to the larger AB 	component observed in $G_2$ compared to $G_1$.  In the semiclassical approximation ($k_f l_e \gg 1$ ) it is possible to  express the   conductance  of a phase coherent ring in terms of interference between scattering amplitudes   of electronic waves, $G = {\rm Re}\left[ \Sigma_{i,j} A_i A_j\exp i\phi_{ij}(B)\right]$, where  indices $i$ and $j$ run on all pairs of possible diffusive trajectories going from one terminal to the other   and $\phi_{ij}(B)= \phi_{i}-\phi_{j}+2\pi B S_{ij}/ \Phi_0 $.  The phase  at zero field $\phi_i$ is the integral $(1/\hbar)\int_i U_{dis}(\vec{r(t)})dt $ on the diffusive trajectory $i$  assuming that the screened  disordered potential $U_{dis}$ varies  smoothly on the  electron Fermi wavelength which is a reasonable assumption in a 2DEG.  $S_{ij}$ is the surface comprised between trajectories $i$ and $j$. Onsager symmetry rules imply that the  2 terminal magneto-conductance  takes the following  form at zero temperature:  
\begin{equation}
G_ =  \Sigma_{i,j} A_i A_j \cos(\phi_{ij})cos(2\pi B S_{ij}/ \Phi_0)
\end{equation}
In the presence of a current through the sample associated to a potential drop $V$, the local electronic density  and consequently the scattering potential  is modified with a term $ d U_{dis} (\vec {r},V)$ which induces  phase shifts, $d \phi_{ij} =\int_i -\int_j (1/\hbar)  d U_{dis}( \vec{r(t)})dt $.  The quantity $G_2 =(dG(V)/dV)_{V=0} $ is then directly related to these phase shifts through:
  
\begin{equation}
G_2  = \Sigma _{i,j} A_i A_j (d\phi_{ij}/dV)\sin(\phi_{ij})\cos(2\pi B S_{ij}/ \Phi_0)
\label{eqg2}
\end{equation}
 Since $d \phi_{ij}$ increases with the length of the interfering trajectories $i$ and $j$,   long trajectories encircling the ring (AB oscillations)   contribute more than trajectories within the same branch of the ring (UCF). This explains the larger relative amplitude of the AB oscillations   and the larger harmonics content in $G_2$ compared to $G_1$ (see  Fig.2 a,b). In a diffusive open system   $d U_{dis}$  is of the order of $eV$ and the main contribution to the typical value of $d \phi_{ij}$ is  $eV\tau_D/\hbar$  and is independent of B leads to the expression of $\delta G_2^S$ of Eq.\ref{astheo}.  This contribution is strongly attenuated   for a   ballistic system with strong interactions, like in \cite{polianski}, where most of the potential drop takes place at the contacts.   Other contributions to   $\delta G_2$  of the order of 1/g  are obtained by taking into account   mesoscopic non local  field dependent fluctuations  in  the  potential $d U_{dis}( \vec{r},B)$, with   symmetric and antisymmetric  components in B,  both of  order of   $ \gamma_{int}eV \delta G_1/g$ at high flux compared to $\Phi_c$.   The antisymmetric component in $d U_{dis}( \vec{r},B)$ is expected to  typically vary like $\gamma_{int} (\delta G_1/g)V\Phi/\Phi_c $  at low flux. We can then deduce  from Eq.\ref{eqg2} the main contributions to the AB oscillating part of $\tilde G_2^{AS}$, which originates from  pairs of  trajectories  encircling the rings: 
\begin{equation}
\tilde G_2^{AS} = \frac {\delta G_1}{ g\hbar} \gamma_{int} \Sigma _{i,j} e A_i A_j  \tau_{ij} \sin(\phi_{ij})\frac{\Phi}{\Phi_c}\cos(2\pi \Phi /\Phi_0)
\end{equation}

There is also a smaller contribution of terms in $ \sin(2\pi\Phi/ \Phi_0)$ of the order of $\delta\tilde  G_1/\delta G_1$ which is less than 0.2 in our experiments. This provides an explanation for  the linear increase of  the amplitude of the AB oscillations of $G_2^{AS}$ in   a  flux range  $\Phi_c$ for rings  with $g>1$ as shown in  Fig.4b. However this effect may not exist in very narrow  rings   \cite{leturq} where conductance fluctuations are not observed. It is interesting that this simple heuristic model can also explain  the  larger value of  $G_2^S$  in a diffusive compared to a ballistic system and the different values  of the ratio $G_2^{AS}/G_2^S$ of the order of  $\gamma_{int}/g$,  respectively 1, in a diffusive, respectively ballistic, system. 
 
In conclusion we have shown evidence of a field asymmetry on the second order response of GaAs/GaAlAs rings  of mesoscopic origin which contains both AB oscillations and conductance fluctuations.   This asymmetry  is characterized by  $\delta G_2^{AS}/\delta G_2^{S}$ and analyzed within theoretical predictions  expressing   this ratio with      only 2 parameters,  the dimensionless conductance of the rings and  the interaction constant   whose value can be determined $\gamma_{int}= 0.90 \pm 0.05$. We have also found  that the relative amplitude of the AB oscillations compared to the UCF is much larger in $G_2$ than in  $G_1$ with the existence of a linear low field modulation in the AB oscillations in the antisymmetric component of $G_2$. These effects  can be understood within a simple   semi-classical description of quantum interference. 
 We thank D. Mailly and F. Pierre for their helpful contribution to the fabrication of the samples as well as  M. B\"uttiker, M.Ferrier, G. Montambaux, B.Reulet, B. Spivak and C. Texier for fruitful discussions. 
{\it Note added} During completion of this work, related experimental work on  gated  quantum dots \cite{marcus} and  small Aharonov Bohm  rings \cite{leturq} appeared.


\begin{thebibliography}{99}
\bibitem{orderofmag} We note that for  a metallic sample $\Delta$ is very small and  $\gamma_{int}$ is very close to 1.
\bibitem{notemarcus}In this paper  it seems that no symmetric component in $G_2$ could be detected.
\end{thebibliography}
\end{document}